\documentclass[aps,reprint,notitlepage,superscriptaddress,showpacs,showkeys]{revtex4-1}
\usepackage{latexsym,amssymb,amsfonts,mathrsfs,amsmath,bm}
\usepackage{graphicx}
\usepackage[usenames,dvipsnames]{xcolor}
\usepackage{soul}
\usepackage[linktocpage,colorlinks=true,linkcolor=blue,citecolor=blue,breaklinks=true,urlcolor=blue]{hyperref}

\renewcommand{\vec}[1]{\bm{#1}}%

\newcommand{\persec}{\text{s}^{-1}}


\begin{document}

\title{Oscillatory surface rheotaxis of swimming \textit{E.~coli} bacteria}  

\author{Arnold J. T. M. Mathijssen}
\affiliation{Department of Bioengineering, Stanford University, 443 Via Ortega, Stanford, CA 94305, USA}
\affiliation{Rudolf Peierls Centre for Theoretical Physics, University of Oxford, 1 Keble Road, OX1 3NP, UK}

\author{Nuris Figueroa-Morales{$^\dagger$}}
\affiliation{Physique et M\'ecanique des Milieux H\'et\`erog\`enes, PMMH, ESPCI Paris, PSL University, CNRS, Sorbonne Universit\'e, Univ Paris Diderot, 10, rue Vauquelin, 75005 Paris, France}
\altaffiliation[Present address: ]{Department of Biomedical Engineering, Penn State University, 508 Wartik Lab, PA 16802, USA}

\author{Gaspard Junot}
\affiliation{Physique et M\'ecanique des Milieux H\'et\`erog\`enes, PMMH, ESPCI Paris, PSL University, CNRS, Sorbonne Universit\'e, Univ Paris Diderot, 10, rue Vauquelin, 75005 Paris, France}

\author{\'Eric Cl\'ement}
\affiliation{Physique et M\'ecanique des Milieux H\'et\`erog\`enes, PMMH, ESPCI Paris, PSL University, CNRS, Sorbonne Universit\'e, Univ Paris Diderot, 10, rue Vauquelin, 75005 Paris, France}

\author{Anke Lindner}
\email{anke.lindner@espci.fr}
\affiliation{Physique et M\'ecanique des Milieux H\'et\`erog\`enes, PMMH, ESPCI Paris, PSL University, CNRS, Sorbonne Universit\'e, Univ Paris Diderot, 10, rue Vauquelin, 75005 Paris, France}

\author{Andreas Z\"{o}ttl}
\affiliation{Rudolf Peierls Centre for Theoretical Physics, University of Oxford, 1 Keble Road, OX1 3NP, UK}
\affiliation{Physique et M\'ecanique des Milieux H\'et\`erog\`enes, PMMH, ESPCI Paris, PSL University, CNRS, Sorbonne Universit\'e, Univ Paris Diderot, 10, rue Vauquelin, 75005 Paris, France}

\date{\today}

\begin{abstract}
Bacterial contamination of biological conducts, catheters or water resources is a major threat to public health and can be amplified by the ability of bacteria to swim upstream.
The mechanisms of this `rheotaxis', the reorientation with respect to flow gradients, often in complex and confined environments, are still poorly understood.
Here, we follow individual \textit{E.~coli} bacteria swimming at surfaces under shear flow with two complementary experimental assays, based on 3D Lagrangian tracking and fluorescent flagellar labelling and we develop a theoretical model for their rheotactic motion. 
Three transitions are identified with increasing shear rate: 
Above a first critical shear rate, bacteria shift to swimming upstream.
After a second threshold, we report the discovery of an oscillatory rheotaxis. 
Beyond a third transition, we further observe coexistence of rheotaxis along the positive and negative vorticity directions. 
A full theoretical analysis explains these regimes and predicts the corresponding critical shear rates. 
The predicted transitions as well as the oscillation dynamics are in good agreement with experimental observations.
Our results shed new light on bacterial transport and reveal new strategies for contamination prevention.
\end{abstract}

\maketitle


\begin{figure*}[t]
\includegraphics[width= \textwidth]{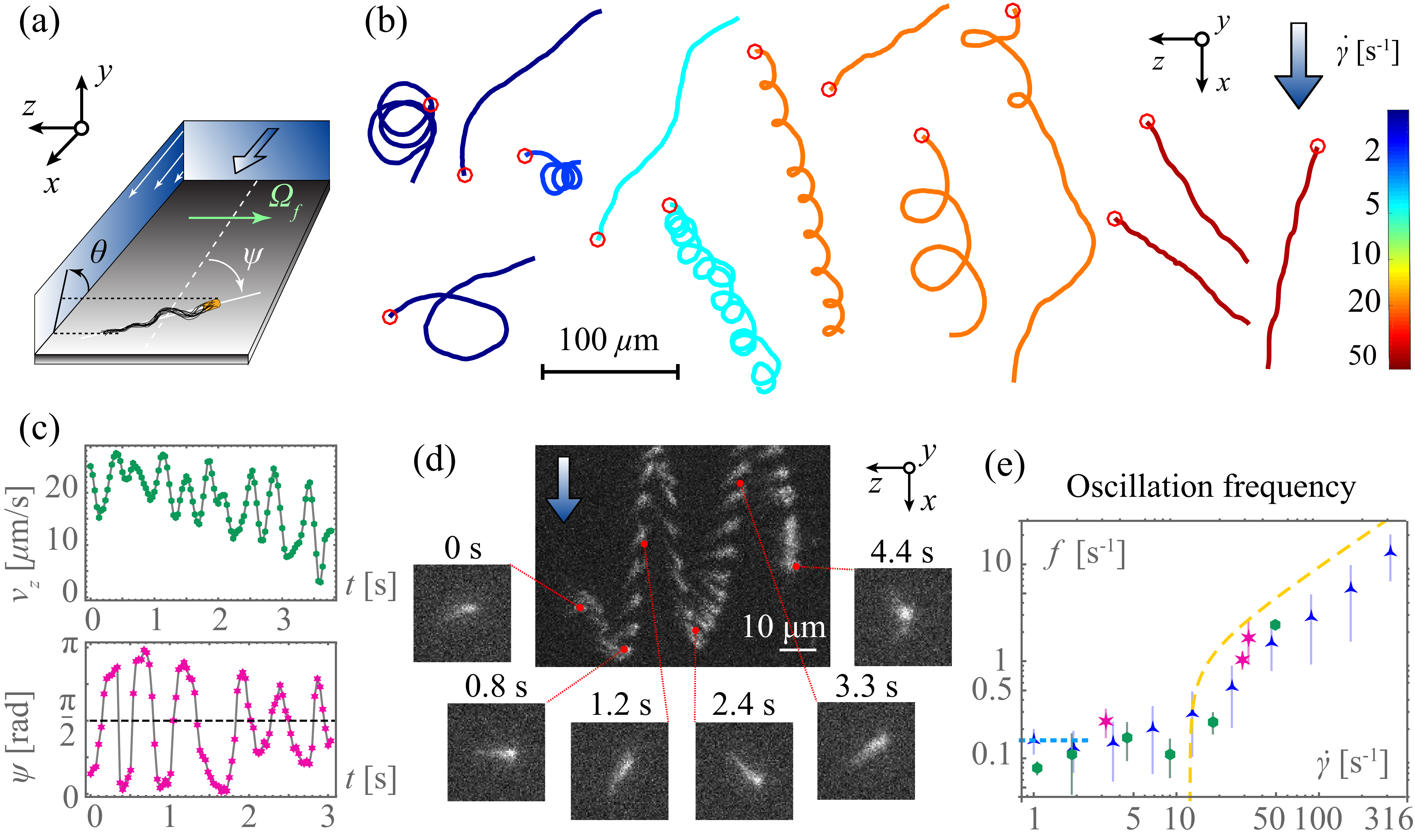}
 	\caption{\label{Fig:1} 
	Experimental observations of oscillatory rheotaxis. 
 	(a) Set-up geometry.
 	(b) Various types of surface trajectories obtained from 3D tracking at shear rates $\dot{\gamma} = 1-50 \text{s}^{-1}$ (colours), shown in the lab frame and arranged according to increasing shear. Circles indicate the initial positions.
 	(c-e) Oscillatory rheotaxis.
 	(c)  Typical temporal evolution of the transverse velocity $v_z(t)$  from a 3D tracking experiment at $\dot{\gamma}= 49 \text{s}^{-1}$, and the in-plane angle $\psi(t)$ from a fluorescence experiment at $\dot{\gamma}= 33 \text{s}^{-1}$. 
 	(d) Time lapse of an oscillating bacterium with fluorescently stained flagella, using 10 fps snapshots overlaid to highlight its trajectory, taken in the Lagrangian reference frame of the average downstream bacterial velocity.
	(e) Oscillation frequency versus shear rate, obtained from Fourier transformation of $v_z(t)$ in 3D tracking experiments (green hexagons), of $\psi(t)$ in fluorescence experiments (magenta stars), and of $\psi(t)$ in simulations (blue triangles). Overlaid are theoretical estimates (Eq.~\ref{eigenvalues}; dashed yellow line) and the circling frequency, $ \Omega_\psi^W/2\pi$, (dotted blue line).
}
 \end{figure*}

Swimming microorganisms must respond to flows in highly diverse and complex environments, at scales ranging from open oceans to narrow capillaries \cite{Bechinger2016, Zoettl2016,mathijssen2018universal}.
To succeed in such diverse conditions, microbial transport often features surprising dynamics.
Microswimmers can accumulate in shear flows \cite{Kessler1985, Durham2009, Rusconi2014, Barry2015} or behind physical obstacles \cite{mino2018coli}, exhibit oscillatory trajectories and upstream motion in Poiseuille flows \cite{Rusconi2014, Garcia2013}, align resonantly in oscillatory flows \cite{Hope2016}, and feature instabilities during rapid expansion \cite{sokolov2016rapid, sokolov2018instability}.
Some of these observations were explained individually by accounting for hydrodynamics, activity and the swimmers' complex shape \cite{Rusconi2014, Ezhilan2015, Zoettl2013, Zoettl2012, Garcia2013, Ezhilan2015, marcos2009separation, Mathijssen2016}.
Altogether, however, the interplay of these non-linear properties is far from trivial and remains largely unexplored.

Moreover, the understanding of surface locomotion is of particular importance due to boundary accumulation \cite{Rothschild1963, Berke2008, Li2009}, but in the presence of walls these dynamics become increasingly intricate \cite{Molaei2014failed, shum2015hydrodynamic, sipos2015trapping, Mathijssen2016hydrodynamics, Mathijssen2018self}.
In quiescent liquids micro-swimmers move in circles \cite{Ward1985, Berg1990, Luzio2005}, but in currents they can orient with respect to gradients in the flow velocity -- an effect called ``rheotaxis'' \cite{Bretherton1961}.
In particular, organisms can reorient to migrate upstream, as observed for sperm cells \cite{Bretherton1961, Kantsler2014, Tung2015}, for \textit{E.~coli} bacteria \cite{Kaya2012, Hill2007, Figueroa-Morales2015, Altshuler2013} and artificial microswimmers \cite{Palacci2015}.
This upstream motion has been analysed theoretically  \cite{Nash2010, Costanzo2012, Uspal2015} and is generally attributed to fore-aft asymmetry of the swimmer shape.
A second type of rheotaxis, at higher flow rates, can reorient organisms towards the vorticity direction \cite{Hill2007, Kaya2012, Marcos2012, Figueroa-Morales2015}, which is attributed to the inherent flagellar chirality \cite{marcos2009separation}.
To date, predictions for this transition from upstream to transversal rheotaxis for bacteria are actively sought after.
Moreover, bacterial rheotaxis at surfaces has been quantified by measuring instantaneous orientation distributions \cite{Kaya2012} or average transport velocities \cite{Figueroa-Morales2015}, but a dynamical picture of the underlying mechanisms is still missing.

Here, we investigate, for the first time, the time-resolved orientation dynamics of \textit{E.coli} bacteria, as a function of applied shear close to walls. 
Two recent experimental techniques are combined with Brownian dynamics simulations and a thorough theoretical analysis. 
In particular, with increasing flow, we identify four regimes separated by critical shear rates:
(\textit{I}) the well-known circular swimming;
(\textit{II}) direct upstream swimming without circling and without oscillations;
(\textit{III}) a novel oscillatory motion, biased towards the direction of positive vorticity;
(\textit{IV}) coexistence of oscillatory swimming to the positive and negative vorticity directions, with dynamical switching between these states.
By monitoring the bacteria with 3D Lagrangian tracking we examine these regimes as a function of the shear rate.
In a second assay the bacterial flagella are stained fluorescently to explicitly survey cellular orientation dynamics and the oscillation frequencies.
Matching these experiments, we model the bacterial rheotaxis by accounting for the cells' chiral nature, hydrodynamic and steric interactions with surfaces, elongation, fore-aft asymmetry and activity. Starting from these individual swimmer-surface-flow interactions we build up our understanding to the observed motility. 
We assess the relative importance of these contributions, and hence explain the full dynamics.
These findings provide a broad understanding of microbial swimming in confined flows and allow to raise suggestions for optimising flow geometries as for example antibacterial channel design.

      \section*{Results}
      
      \subsection*{Experimental observations}

 \begin{figure*}[t]
\includegraphics[width=\textwidth]{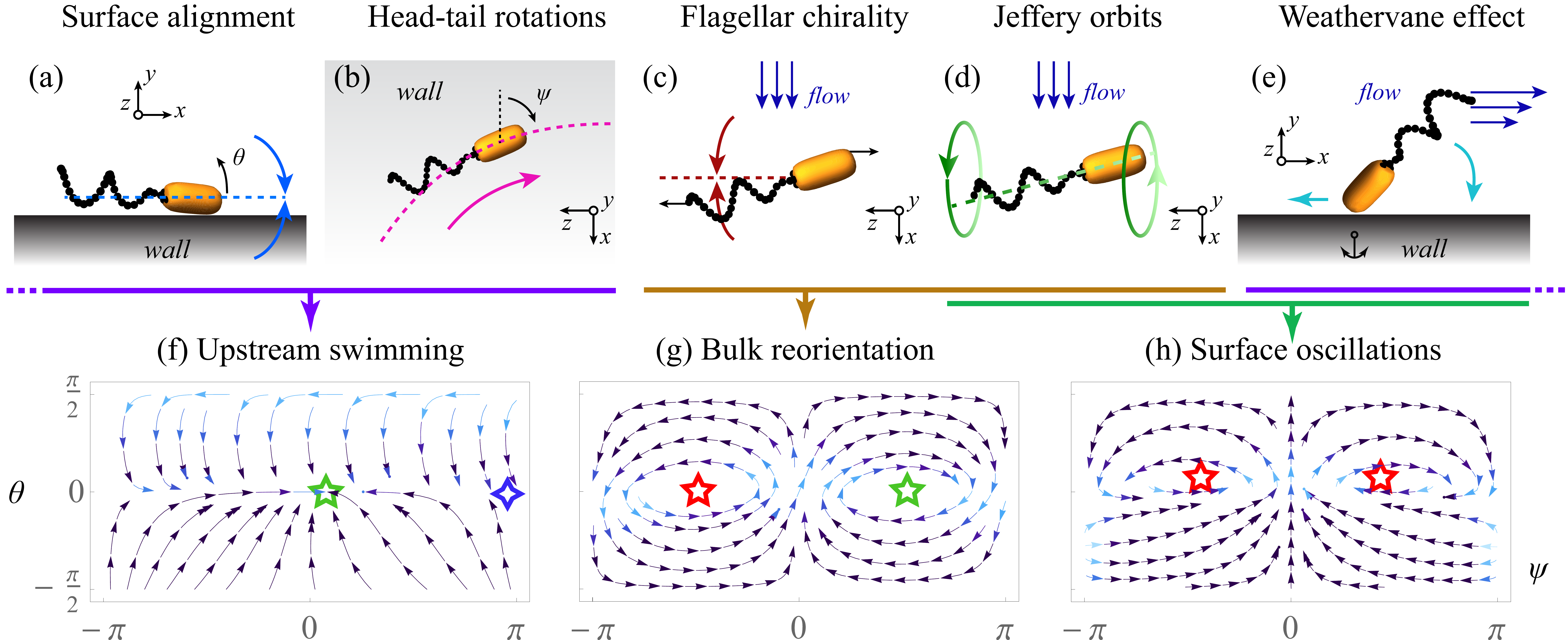}
\caption{\label{Fig:2}
  Summary of  reorientation mechanisms included in the model.
\textit{Wall effects}:
(a) Steric and hydrodynamics interactions align swimmers with surfaces.
(b) Clock-wise torque from the counter-rotation of the cell body and flagella. 	
\textit{Flow effects}: 
(c) Left-handed helical flagellar bundle in shear reorients swimmers to the right.
(d) Jeffery orbits of elongated bacteria. 	
\textit{Flow-wall coupling}: 	
(e) Weathervane effect reorients swimmers to the upstream direction.
The corresponding orientation dynamics for (a)-(e) in $\psi-\theta$ phase space are shown in Supplementary Fig.~S1. 
Combinations of the different effects give
(f) Swimming in the upstream direction [a,b,e].
(g) Bulk reorientation, biased to the right due to flagellar chirality [c,d].
(h) Oscillatory swimming, oriented slightly upstream due to the weathervane effect [d,e].
Green (red) stars are stable (unstable) orientation fixed points, and the blue diamond is a saddle point.
The parameters used are given in the caption of Fig.~\ref{Fig:3}, with shear rate $\dot{\gamma} = 10 \persec$.
}
\end{figure*}

We observe the dynamics of \textit{E.~coli} bacteria at surfaces under flow [Fig.~\ref{Fig:1}(a)].
Two independent experimental realisations are used.
First, we employ a novel 3D tracking technique \cite{darnige2017lagrangian} that provides full 3D trajectories of swimming bacteria over large distances, revealing their long-time Lagrangian dynamics. 
The bacteria used in these experiments are smooth swimmers, \textit{E.~coli} strain CR20, a mutant that almost never tumbles and moves with a typical swimming speed of $v_s = 26\pm4\mu$m/s. 
The experimental device is a rectangular channel made in PDMS and a constant flow is applied with a syringe pump. 
The channel height is $H = 100 \mu$m, the width $W = 600\mu$m and its length is of several millimetres. 
Hence, the shear rate at the bottom wall is defined as $\dot{\gamma}=4 V_\text{max}/H$, where $V_\text{max}$ is the maximum flow velocity at the channel centreline.
Bacterial trajectories are only selected when they are located more than 100~$\mu$m from the lateral side walls and less than 5~$\mu$m from the bottom surface, so that the wall  shear rate is constant 
and 3D trajectories are nearly identical to the $x$-$z$ projections [see Materials \& Methods (MM) \S\ref{subsec:ExperimentalProtocols} for details].

Typical 3D trajectories for shear rates $\dot{\gamma}= 1-50 \persec$ are displayed in Fig.~\ref{Fig:1}(b), in the laboratory frame. 
With increasing shear, we observe a range of different dynamics.
Interestingly, at small shear rates the well-known circular motion \cite{Ward1985, Berg1990, Luzio2005} starts to evolve towards cycloid motion with a bias ``to the right''.
Here we define the term ``to the right'' as the direction of the vorticity vector, $\vec{\Omega}_f = -\dot{\gamma} \hat{\vec{z}}$ [Fig.~\ref{Fig:1}(a); green arrow].
Subsequently, circles become suppressed and, instead, upstream motion is observed. 
When further increasing the shear rate bacteria are transported downstream more strongly and the laboratory frame trajectories bend into the direction of the flow. 
These trajectories are mostly oriented towards the right, as reported previously \cite{Kaya2012}, but for the first time we observe that swimming towards the left can also occur. 
Note that different types of trajectories may coexist, likely due to variations in bacterial shape, the distance from the wall and other sources of noise inherent to living bacteria.

Surprisingly, an oscillatory motion appears in these trajectories at frequencies very different from the flagellar and body rotation.
These undulations are visible in the trajectories at the highest shear rate and can be identified clearly by looking at $v_z(t)$, the velocity component transverse to the flow direction [Fig.~\ref{Fig:1}(c); top panel].
However, since the 3D tracking technique does not provide direct access to bacterial orientation, we also perform a second and complementary set of experiments.
Here we use a genetically modified strain of bacteria, from the AB1157 wild-type (AD1)~\cite{schwarz-linek2016escherichia}, with a fluorescently labelled body and flagella so that the cell orientation is directly visualised.
This wild-type strain can tumble, but we only select trajectories without tumbles, which can be easily identified from the images.
The channel dimensions are $H = 20 \mu$m and $W = 200 \mu$m, and again bacterial dynamics are captured only within a maximal distance of $5~\mu$m from the bottom surface.
In this strong confinement high shear rates can be obtained using relatively small flow velocities, which facilitates straightforward manual tracking, but at the cost of a more variable shear rate in $y$ compared to the 3D tracking.
These fluorescence experiments unambiguously demonstrate the existence of oscillatory motion around a stable position [Fig.~\ref{Fig:1}(d)], shown here for an example swimming to the right.
Moreover, they provide an immediate measure of the orientation angle dynamics $\psi(t)$ [Fig.~\ref{Fig:1}(c); bottom panel]. 

To quantify this oscillatory rheotaxis, for the 3D tracking and fluorescence assays, we extract the oscillation frequencies from Fourier transformation of $v_z(t)$ and $\psi(t)$ respectively [see MM \S\ref{subsec:ExperimentalProtocols}c].
In both experiments the measured frequencies indicate a cross-over [Fig.~\ref{Fig:1}(e)].
At small shear rates we find a constant frequency, corresponding to the circular swimming, and after a certain shear ($\dot{\gamma} \approx 10 \persec$) we observe an increase of the frequency corresponding to the oscillatory trajectories. 
Note, oscillatory rheotaxis should not be confused with the wobbling dynamics due to flagellar rotation \cite{Bianchi2017}, which have much higher frequencies and are distinctly different.

In the next sections we will develop a comprehensive model that explains these complex dynamics and predicts the corresponding oscillation frequencies.

 \subsection*{Theoretical building blocks}

In order to understand the rich behaviour of our experimental findings, we first identify and summarise the individual mechanisms that affect bacterial orientations. 
We distinguish between wall effects, flow effects, and the coupling of these. 
In the next sections we then combine these building blocks and describe their non-trivial interplay.

We model a bacterium consisting of an elongated body and a left-handed flagellar bundle, subject to shear flow at a surface [Fig.~\ref{Fig:2}].
We explicitly model both the in-plane angle $\psi \in \{-\pi,\pi\}$ and the pitch (i.e. dipping) angle $\theta \in \{-\pi/2,\pi/2\}$ [Fig.~\ref{Fig:1}(a)].
Note, the bacterial conformation in principle also depends on the flagellar helix phase angle, which can lead to phase-dependent wobbling motion \cite{Hyon2012, Bianchi2017}, but owing to its fast flagellar rotation this angle is averaged over.
The orientation of a swimmer at the surface then evolves as 
\begin{equation}
\dot{\psi} = \Omega_\psi(\psi,\theta), \qquad
\dot{\theta} = \Omega_\theta(\psi,\theta),
\label{Eq:dPdT}
\end{equation}
where the reorientation rates $\Omega_\psi$ and $\Omega_\theta$  stem from three main contributions,  $\Omega = \Omega^W + \Omega^F + \Omega^{V}$, that account for the presence of the wall ($\Omega^W$), local shear flow ($\Omega^F$), and surface-flow coupled effects ($\Omega^{V}$).

\paragraph{Wall effects.} 
First, in the absence of flow, hydrodynamic swimmer-wall interactions \cite{Berke2008, Spagnolie2012} and steric interactions \cite{Li2009} enable bacteria to swim at a stable orientation approximately parallel to the wall \cite{Lauga2006, Li2008, dunstan2012two, Bianchi2017} [Fig.~\ref{Fig:2}(a)].
We model this surface alignment as
$\Omega_\theta^W(\theta)  =- \nu_{W}\sin 2 (\theta-\theta_0) \left(1 + \frac G 2 (1 + \cos^2 \theta)\right)$ 
where $G= \frac{\Gamma^2-1}{\Gamma^2+1} \lesssim 1$ is a geometric factor describing the elongation of the bacterium with effective aspect ratio $\Gamma$. The prefactor $\nu_{W}$ is an effective angular rate capturing both the hydrodynamic and steric contributions, and the equilibrium pitch angle $\theta_0 < 0$ represents the fact that bacteria on average assume a small angle pointing towards the wall \cite{Bianchi2017}. 
For simplicity, we set $\theta_0=0$ but we verified that non-zero values do not alter our results qualitatively. Note, we still explicitly model dynamics in $\theta$.

The second wall effect stems from the counter-rotation of the bacterial head and flagellar bundle.
Near solid surfaces this leads to a hydrodynamic torque leading to circular motion in the clockwise direction \cite{Luzio2005, Lauga2006} [Fig.~\ref{Fig:2}(b)].
The associated reorientation rate in the $\psi$ direction is approximated by 
$\Omega_\psi^W(\theta) = \nu_{C} (1-3\sin^2\theta + G \cos^2\theta(1+3\sin^2\theta) $
\cite{Spagnolie2012}, 
which simplifies to a constant rotation $\Omega_\psi^W \approx \nu_{C} (1+G)$ for small $\theta$.

\paragraph{Flow effects.}
Second, we discuss the contributions due to shear flows.
Elongated objects such as rods and fibres, or dead bacteria \cite{Kaya2009}, perform Jeffery orbits \cite{Jeffery1922} such that the orientation vector performs a periodic motion about the vorticity ($z$) direction [Fig.~\ref{Fig:2}(d)], given by
$\Omega_\psi^J =   \frac{\dot{\gamma}}{2} (1+G) \sin\psi\tan\theta$
and
$\Omega_\theta^J =  \frac{\dot{\gamma}}{2} (1 - G \cos 2 \theta) \cos\psi$.
In the presence of walls the orbit amplitudes decay because of the surface alignment [see previous paragraph] but their reorientation rate (frequency) is not affected significantly, as simulated in detail for passive ellipsoidal particles \cite{Pozrikidis2005}.

The second flow effect stems from the chirality of the bacterial flagella, making cells reorient towards the vorticity direction \cite{Makino2005, Marcos2009}.
Together with activity this enables stream-line crossing, which in the bulk leads to a net migration of bacteria ``to the right'' \cite{Marcos2012}. 
We compute this effect using resistive force theory (RFT) applied to a helical flagellar bundle under shear flow, extending the calculations in Ref.~\cite{Marcos2012} for all body cell orientations [see MM \S\ref{subsec:RTFcalculation} and Fig.~\ref{Fig:2}(c)].
This yields the chirality-induced reorientation rates
\begin{equation}
\begin{split}
\Omega_\psi^{H} &=   \dot{\gamma} \bar{\nu}_{H} \cos\psi \frac{\cos 2 \theta }{\cos\theta}, \quad
\Omega_\theta^{H} =  \dot{\gamma} \bar{\nu}_{H} \sin\psi\sin \theta.
\label{Eq:OmH}
\end{split}
\end{equation}
with a prefactor $\bar{\nu}_{H} \ll 1$ that solely depends on the geometry of the bacterium.

\paragraph{Weathervane effect due to wall-flow coupling.}
Third, we introduce a term that has been identified as an important contribution for sperm rheotaxis \cite{Tung2015, Kantsler2014}.
The swimmer body experiences an effective \textit{anchoring} to the surface when pointing towards it, because its hydrodynamic friction with the wall is larger than that of the flagellar bundle \cite{Daddi2018state}, an effect explained by lubrication theory \cite{Tung2015}.
Consequently, the flagella are advected with the flow, like a \textit{weathervane}.
Then, the bacterium orients upstream \cite{Miki2013, Kantsler2014, Tung2015}, which we model using the reorientation rates
\begin{equation}
\Omega_\alpha^{V} = - \dot{\gamma} \bar{\nu}_{V} \sin( \alpha )
 \left[\frac 1 2\left(1 - \tanh \frac{\theta}{\theta_{V}} \right) \right], 
\label{Eq:OmWV}
\end{equation}
for both $\alpha = \{ \theta, \psi \}$.
The hyperbolic tangent, with a constant $\theta_{V}$ depending on the cell geometry, accounts for the fact that the asymmetry in friction reduces when the swimmer faces away from the surface, $\theta >0$, where the weathervane effect disappears [Fig.~\ref{Fig:2}(e)].
This notion was not included in the single-angle descriptions used for sperm \cite{Kantsler2014, Tung2015}.

\subsection*{Combining the individual reorientation mechanisms}

\begin{figure*}[t]
\includegraphics[width= 0.9 \textwidth]{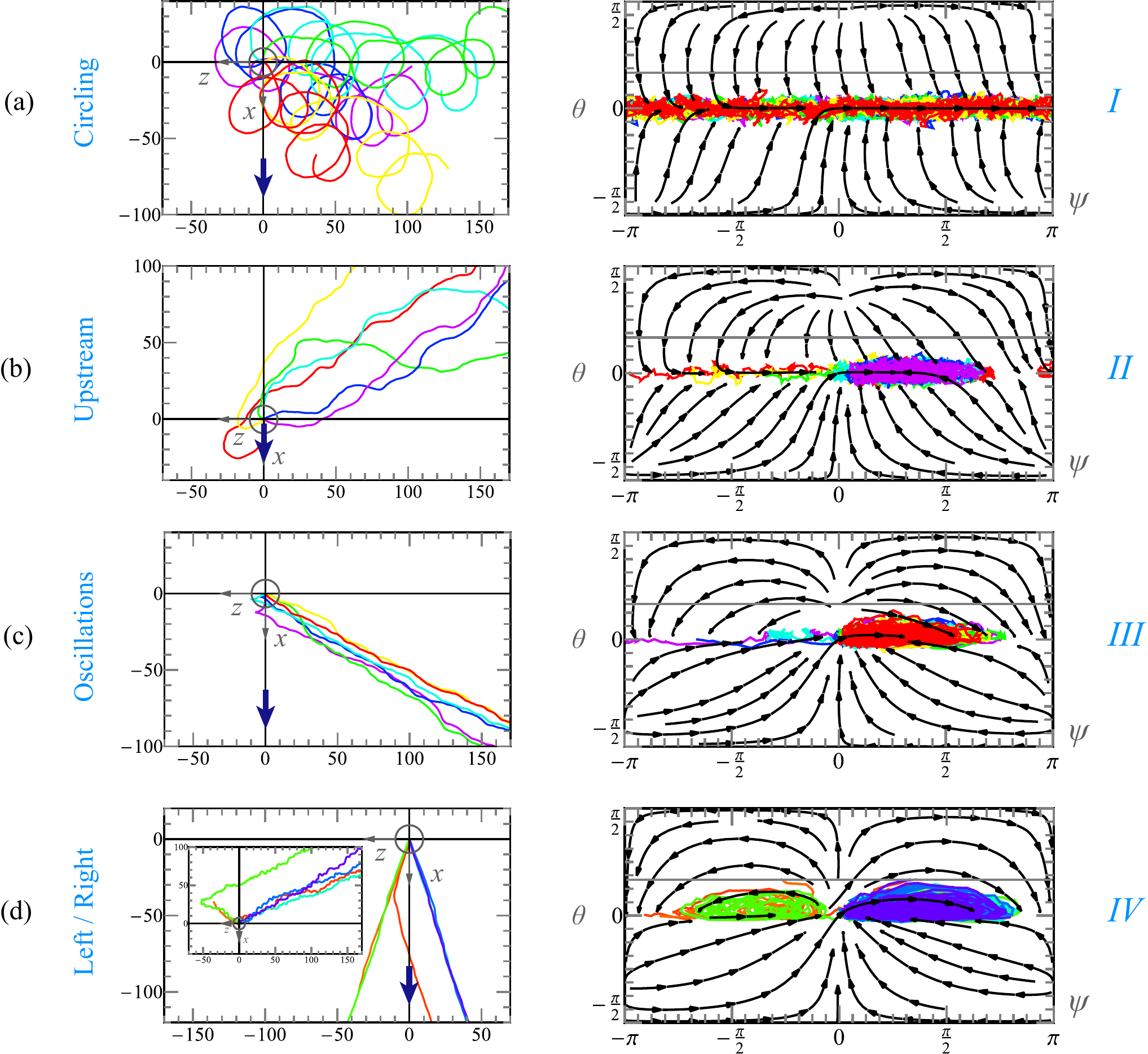}
 \caption{\label{Fig:3}
Simulated trajectories of surface rheotaxis (upper panels) in the laboratory frame and the corresponding orientation dynamics (lower panels) in the four different regimes with increasing shear rate: 	
 	(I, $\dot{\gamma} =1 \persec$)  Circular swimming with a bias to the right. 	
 	(II, $\dot{\gamma} =5 \persec$)  Upstream motion. 	
 	(III, $\dot{\gamma} =20 \persec$)  Oscillatory motion, increasingly more to the right.
(IV, $\dot{\gamma} =50 \persec$) Coexistence between swimming to the right and to the left, with dynamical switching between these. The inset show dynamics in the frame co-moving with the flow.
Grey circles indicate the initial swimmer positions.
Parameters:
$\Gamma=5$,
$\nu_{W}=3 \persec$,
$\nu_{D}=0.5 \persec$,
$\bar{\nu}_{H}=0.02$,
$\bar{\nu}_{V}=0.75$,
$\theta_{0} = 0$,
$\theta_{V} = 0.04$,
$\theta_e = \pi/6$,
$v_s = 20 \mu$m$\persec$,
$h_s = 0.5 \mu$m,
$D_r=0.057\persec$.
}
 \end{figure*}

Having described the individual effects of the wall and the flow on bacterial reorientation, we can now combine these terms to begin to understand more complex dynamics. 
First of all, by joining the contributions from surface alignment and head-tail rotations, we recover the well-known circular swimming \cite{Luzio2005, Lauga2006}.
However, when we also add the weathervane effect (\ref{Eq:OmWV}) the cells break out of the circular kinetics and swim upstream, which corresponds to a stable fixed point in their orientation space [Fig.~\ref{Fig:2}(f)]. 
This Adler transition has also been observed for sperm cells \cite{Tung2015}.

Second, combining the effects of Jeffery orbits and chirality (\ref{Eq:OmH}), we recover bulk rheotaxis \cite{Marcos2012}.
Recast into the language of dynamical systems, the symmetry breaking leading to preferred motion `to the right' can be classified as a stable `spiral' fixed point in $\psi-\theta$ phase space [Fig.~\ref{Fig:2}(g)].

Third, merging the Jeffery orbits in the bulk and the weathervane effect (\ref{Eq:OmWV}) for cells near a surface, we find that Jeffery's periodic motion about the vorticity directions ($\pm\hat{\vec{z}}$) now shifts to oscillations about a vector pointing more and more upstream. 
This already hints towards the observed oscillatory rheotaxis, but the corresponding fixed points are not stable [Fig.~\ref{Fig:2}(h)]. 
To understand the experimental trajectories accurately, therefore, we must include all terms together and also add fluctuations, as described in the next section.

\subsection*{Brownian dynamics simulations}


\begin{figure*}[t]
\includegraphics[width=\textwidth]{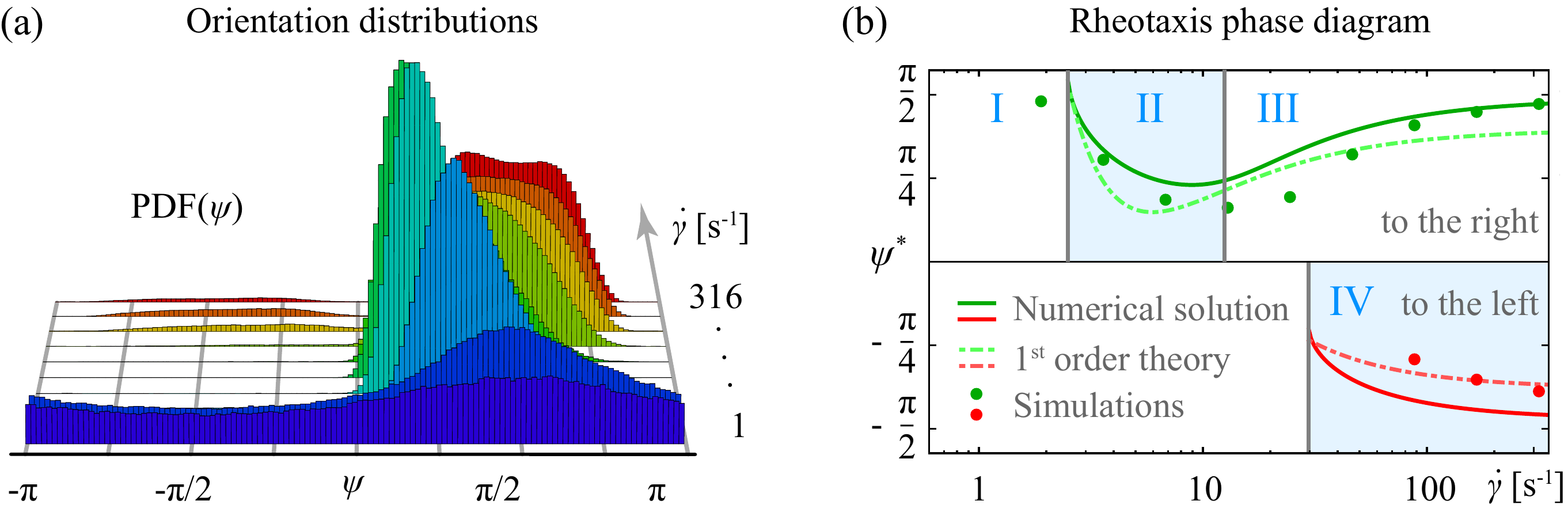} 
\caption{\label{Fig:4}
	Bacterial orientation as a function of applied shear.
	Parameters used are the same as in Fig.~\ref{Fig:3}.
	(a) Distribution of the in-plane angle $\psi$, obtained by simulating $N=10^4$ trajectories over long times until steady state is reached, at each shear rate $\dot \gamma = 10^{2.5 (i-1)/9}$ with $i\in[1,10]$. 
	(b) Equilibrium in-plane angle $\psi^*$, obtained numerically (solid lines), analytically (dashed lines) and in simulations (points), with critical shear rates (grey vertical lines).
}
 \end{figure*}

Combining all the aforementioned contributions, we solve the orientation dynamics by integrating Eqs.~(\ref{Eq:dPdT}) for a range of constant shear rates, together with rotational noise ($D_r=0.057\persec$ \cite{Drescher2011}), as detailed in MM~\S\ref{subsec:RheotaxisSimulations}.
All our parameter values have been estimated carefully from previous experiments and numerical results, as discussed in MM \S\ref{subsec:EstimateParameters}, and the results are qualitatively robust for changes in these parameters.
A simulated trajectory starts with a random $\psi$ and a slightly negative pitch angle, $\theta = -0.1 \pi$, and finishes when it reaches a given escape angle $\theta_e$ \cite{Drescher2011, Schaar2015, Mathijssen2016b}.
Subsequently, the spatial dynamics are found by computing the velocity parallel to the surface, at a constant swimming speed $v_s$ plus the downstream advection with velocity $\vec{v}_f = \dot{\gamma} y \hat{\vec{x}}$ based on the shear rate and the distance from the wall, $y$. 
Hence, Figure~\ref{Fig:3} shows typical trajectories and the orientation dynamics in $\psi$-$\theta$ space for different shear rates $\dot{\gamma}$ and initial conditions.
We identify four regimes ($I$-$IV$) separated by critical shear rates:

At weak flows (regime \textit{I}) the bacteria move in circular trajectories, with a drift to the right [Fig.~\ref{Fig:3}(a)].
Above a critical shear rate, found in the simulations at $\dot{\gamma}_{c_1}^\text{sim} \approx 3 \persec$, they no longer move in circles but swim stable to the right and slightly upstream  (regime \textit{II}) [Fig.~\ref{Fig:3}(b)]. 
This Adler transition \cite{Tung2015} stems from the competition between the constant head-tail rotations and the weathervane effect that increases with flow.
Owing to noise, coexistence between circling and stable swimming may exist close to $\dot{\gamma}_{c_1}$, and oscillations may also appear already, as discussed below.

Above a second critical shear rate, $\dot{\gamma}_{c_2}^\text{sim} \approx 15 \persec$ (regime \textit{III}), an oscillatory motion directed to the right emerges [Fig.~\ref{Fig:3}(c)].
Similar to the first transition, the oscillations arise because the flow contributions now outweigh the surface terms that do not increase with shear.
Particularly the Jeffery and weathervane effects govern the oscillation dynamics, as discussed in the theoretical predictions section. 
A simplified pictorial summary of this oscillation process is provided in Figure~\ref{Fig:5}.
However, the equilibrium angles about which the cells oscillate, $\psi^* \sim \frac \pi 2$ and $\theta^*\sim 0$, still depend strongly on the other terms, as derived below.
Especially the surface alignment is necessary for stability, so in general the dynamics remain a complex interplay between all contributions and fluctuations.

Above a third critical shear rate, $\dot{\gamma}_{c_3}^\text{sim}  \approx 30 \persec$ (regime \textit{IV}), oscillatory swimming to the left arises [Figs.~\ref{Fig:3}(d)], in coexistence with the aforementioned oscillations to the right. 
In phase space, this is defined by the emergence of a stable spiral fixed point, at $\psi^* \sim - \frac \pi 2$ on the left. 
Moreover, bacteria may switch dynamically between the left and right (orange and green trajectories).
However, this mode of rheotaxis is rare as the flagellar chirality term (\ref{Eq:OmH}) gives a bias to the right that also grows with shear.

Throughout these regimes, the orientation distributions have a complex dependence on the shear rate [{Fig.~\ref{Fig:4}(a)].
In the absence of flow, the orientations are uniformly distributed, as expected [dark blue distribution].
In regime \textit{I} the circular swimming is biased to the right, giving a peak in the distribution at $\psi^*\sim \frac \pi 2$ [blue]. 
In regime \textit{II} the swimmers move more and more upstream, so the peak shifts to $\psi^* \gtrsim 0$ [cyan] due to the weathervane effect, as also observed for sperm cells \cite{Tung2015}.
In regimes \textit{II-III} the cells shift from upstream to the right again, $\psi^* \sim \frac \pi 2$ [green], in agreement with previous studies \cite{Hill2007, Kaya2012, Figueroa-Morales2015}. 
This is explained by the weakening of the weathervane effect, because the surface anchoring is reduced by the Jeffery term that tries to rotate the bacterium away from the surface.
Consequently, the optimal shear rate for upstream orientation is found at $\dot{\gamma}_{u}^\text{sim}  \approx 6.8 \persec$, where the in-plane orientation is $\langle \psi_{u}^\text{sim} \rangle  \approx 39.6$ degrees [cyan]. 
Finally, in regime \textit{IV} swimming to the left emerges at $\psi^* \sim \frac \pi 2$ [orange].
In the absence of tumbling, the fraction of bacteria oriented to the left is $\sim 4 \%$ at large shear rates [red], but with tumbles this increases substantially as dynamical switching is enhanced [Supplementary Fig.~S2].

 \begin{figure*}[t]
\includegraphics[width= \textwidth]{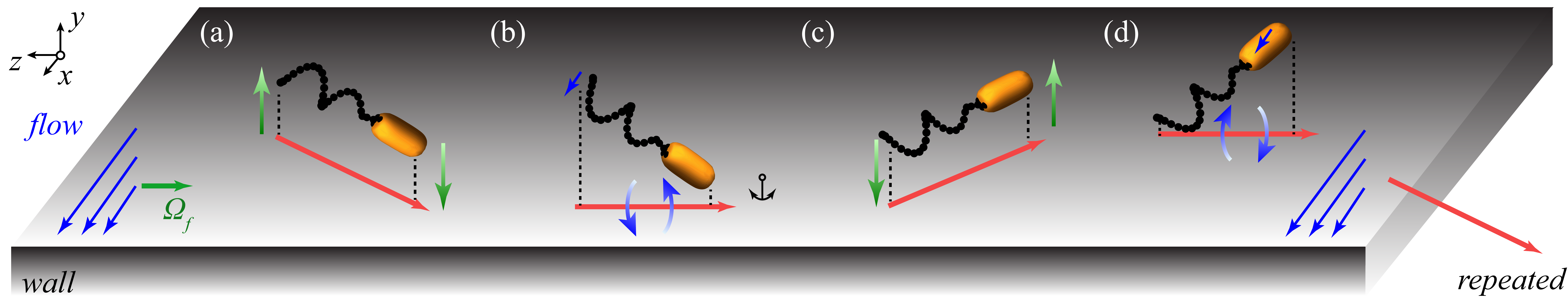} 
\caption{
\label{Fig:5}
Sketch of the oscillatory rheotaxis mechanism. 
Here the bacterium is initially oriented towards the right and slightly downstream, and red arrows show the \textit{projection} of the cell onto the surface.
Then, the oscillations can be envisaged as a 4-step process:
(a) The vorticity pushes the body down into to surface and lifts the flagella up.
(b) Then the flow advects the flagella faster than the body, rotating the bacterium about the $y$ axis to the upstream direction.
The weathervane effect enhances this rotation as the cell pivots about the anchoring point.
(c) Now the vorticity pushes the flagella into to wall and lifts the body up.
(d) Subsequently the body is advected faster, rotating the swimmer back to the downstream direction. 
This cycle is repeated, leading to oscillatory trajectories.
Note that this is a simplified picture and all surface and flow effects [Fig.~\ref{Fig:2}] contribute to the dynamics at any one time.
}
 \end{figure*}

\subsection*{Analytical predictions}

The observed and simulated transitions can also be met with theoretical predictions, as detailed in MM \S\ref{subsec:ModelFixedPoints}. 
We consider the case in the absence of noise, and analyse the dynamics in ($\psi, \theta$) orientation space. 
The equilibrium orientations, $\psi^*$ and $\theta^*$, can be found from the fixed points of Eq.~\ref{Eq:dPdT}.
We obtain the exact values numerically, and also estimate them analytically by linearising Eq.~\ref{Eq:dPdT} about $(\psi, \theta) = (\pm \pi/2,0)$ and solving for the two roots.

The equilibrium orientations found with this analysis are depicted in Fig.~\ref{Fig:4}(b) [solid and dashed lines respectively] as a function of shear rate.
The dynamics to the right (left) are shown in green (red).
Importantly, these deterministic solutions agree well with the simulated dynamics that include noise [Fig.~\ref{Fig:4}(b); points] obtained from the average peak positions of the steady-state distributions [Fig.~\ref{Fig:4}(a)].

Using this framework we also obtain the first critical shear rate,
\begin{align}
\dot{\gamma}_{c_1}^\text{th} =  2(G+1) \nu_{DR} / \bar{\nu}_{V} \approx 2.6 \persec,
\end{align}
and the second critical shear rate, $\dot{\gamma}_{c_2}^\text{th} \approx 13.6 \persec$. 
The third critical shear rate does not follow from linearisation but is found numerically, $\dot{\gamma}_{c_3}^\text{th} \approx 29.7 \persec$.
These are combined to form the complete phase diagram [Fig.~\ref{Fig:4}(b); vertical grey lines].
Moreover, the optimal shear rate for upstream swimming is found to be $\dot{\gamma}_{u}^\text{th} \approx 5.95 \persec$ and the corresponding upstream orientation, $\psi_{u}^\text{th} \approx 26.7$ degrees. 
These four predictions agree very well with the numerical values obtained from our simulations.

Finally, we also derive the oscillation frequency. 
Interestingly, this solution [Fig.~\ref{Fig:1}(e); yellow line] increases from zero after the second critical shear rate.
For high shear rates the frequency can further be approximated as
\small
\begin{align}
\label{Eq:OscFreq}
\omega_O =
\frac{\dot{\gamma}}{4}   \sqrt{4 (1-G) \frac{\bar{\nu}_{V}}{\theta_V} + 4 (1-G^2) - \bar{\nu}_{V}^2 \pm 8 \bar{\nu}_{H} \bar{\nu}_{V} - 16 \bar{\nu}_{H}^2 },
\end{align} 
\normalsize
which increases linearly with the applied shear rate.
This solution offers a good agreement with our simulations and experiments, as discussed next.

	\section*{Discussion}

Our model allows to predict the full time-resolved orientation dynamics of {\it E. coli} close to surfaces as a function of the applied shear rate. 
All types of trajectories predicted by our model have also been observed experimentally [Fig.~\ref{Fig:1}(b)]. 
First, at lower shear rates we see the transition from circular motion to upstream swimming [light blue trajectories]. 
This is when the upstream biasing weathervane effect \cite{Tung2015, Kantsler2014} takes over from the torque due to head-tail rotations.
Second, at intermediate shear rates we see a transition from smooth swimming to oscillations [orange trajectories].
This is when the Jeffery and weathervane effects couple to drive oscillations, and like in bulk rheotaxis \cite{Marcos2012} the chirality of the helix sets a bias to the right. 
Third, at the highest shear rates we also see the switching to left-oriented motion [red trajectories].
Here the left direction becomes stable with respect to fluctuations, although the chirality still drives a main preference to the right. 
Of course, in these observations visual differences can arise from fluctuations, variations in swimming speed and distance with respect to the wall. 
The critical shear rates predicted from both numerical and analytical findings are in reasonable agreement with those observed experimentally. 
In addition, the experimentally observed angular dynamics are very well captured.
In particular, the oscillation frequencies obtained from Fourier transformation of the experimental trajectories [Fig.~\ref{Fig:1}(e); green hexagons and magenta stars] match the simulated frequencies [blue triangles] and the analytical prediction [yellow line] quantitatively. 

Beyond our theory using equilibrium analysis without noise, we expect that fluctuations will contribute to three main effects: 
First, they sustain oscillations despite the wall-alignment damping, as observed in experiments [Fig.~\ref{Fig:1}] and simulations with noise [Fig.~\ref{Fig:3}]. 
Second, they also allow for oscillations to emerge below the critical shear $\dot{\gamma}_{c_2}$, as seen in Fig.~\ref{Fig:3}(b). 
Third, they facilitate dynamical switching between left and right-orientated rheotaxis [Fig.~\ref{Fig:3}(d)], which can be envisaged as jumps in orientation space. 
In our experiments we have used non-tumbling cells to study oscillatory rheotaxis in a controlled manner, and bacterial tumble events will additionally contribute to these oscillations and left-right switches.
Accordingly, in simulations with tumble events we see more swimming to the left at higher shear [Supplementary Fig.~S2].
Moreover, for wild-type strains the average population dynamics at surfaces is also expected to depend quite sensibly on the run-time distribution.

In conclusion, we demonstrate that bacterial surface rheotaxis can be categorised in four regimes, separated by shear-regulated transitions. 
We observe these regimes using a Langrangian 3D cell tracking technique that allows us to follow bacteria in a flow over long times, and with independent measurements by fluorescently staining the cell flagella to monitor the cell orientation explicitly.
A comprehensive model delineates these dynamics by combining previously postulated contributions with newly derived rheotactic terms near a surface.
Simulating this model yields the cellular orientation distributions and their oscillation frequencies with increasing shear rates, and a theoretical analysis of these equations allows to predict the corresponding critical shear rates.
We find that both the bulk rheotaxis term [Eq.~\ref{Eq:OmH}] and the flow-wall coupling [Eq.~\ref{Eq:OmWV}] play significant roles, through a previously under-appreciated dynamical interplay of the bacterial in-plane and pitch angles. 
Even if the bulk rheotaxis prefactor is small, the bias it generates is additive over time and also affects the orientation distribution of bacteria entering the surface in favour of swimming to the right.
Importantly, this bias raises the suggestion that upstream swimming in conventional cylindrical pipes could be deterred with a right-handed surface patterning that spirals inside the duct.
Then right-oriented cells would hit a barrier when swimming upstream, and only a small amount of left-oriented cells could pass if not advected downstream already.
Moreover, oscillations at the surface increase the probability to detach, thus tuning $\dot{\gamma}_{c_2}$ could modify the average residence time on the wall and thus the ability for cells to contaminate upstream areas or initiate biofilms.
Together, these results shed new light on bacterial transport and allow for the development of strategies for controlling surface rheotaxis.

\renewcommand\thefigure{M\arabic{figure}} 
\setcounter{figure}{0}
\renewcommand\theequation{M\arabic{equation}} 
\setcounter{equation}{0}
\renewcommand\thesubsection{\arabic{subsection}} 
\setcounter{subsection}{0}
\renewcommand\thesubsubsection{\alph{subsubsection}} 
\setcounter{subsubsection}{0}

\begin{widetext}
\section*{Methods}
\end{widetext}

\footnotesize

\subsection{Experimental details}
\label{subsec:ExperimentalProtocols}

\subsubsection{3D tracking experiments}

The bacteria are smooth swimmers \textit{E.~coli} (CR20), a mutant strain that almost never tumble. Suspension are prepared using the following protocol: bacteria are inoculated in 5mL of culture medium (M9G: 11.3 g/L M9 salt, 4 g/L glucose, 1 g/L casamino acids, 0.1mM CaCl$_2$,  2mM MgSO$_4$) with antibiotics and grown over night. In this way, bacteria with a fluorescently stained body are obtained. Then the bacteria are transferred in Motility Buffer (MB: 0.1mM EDTA, 0.001mM l-methionine, 10mM sodium lactate, 6.2mM K$_2$HPO$_4$, 3.9mM KH$_2$PO$_4$) and supplemented with L-serine and polyvinyl pyrrolidone (PVP). The addition of L-serine increases the bacteria mobility and PVP is classically used to prevent bacteria from sticking to the surfaces. The interactions that come into play using this system are thus solely steric and hydrodynamic. After incubating for an hour in the medium to obtain a maximal activity, the solution was mixed with Percol (1:1) to avoid bacteria sedimentation. Under these conditions, the average swimming speed is $v_s = 26 \pm 4 \mu$m/s. 
For the experiments the suspension is diluted strongly such as to be able to observe single bacteria trajectories without interactions between bacteria.

The experimental cell is a rectangular channel made in PDMS using soft lithography techniques. The channel height is $h = 100 \mu$m, the width $w = 600\mu$m and its length is of several millimeters. 
Using a syringe pump (dosing unit: Low Pressure Syringe Pump neMESYS 290N and base: Module BASE 120N) we flow the suspension inside the channel at different flow rates (1 1,88 4,5 9 18 50 nL.s$^{-1}$), corresponding to wall shear rates of 1-50 s$^{-1}$.
To have access to the 3D trajectories of single bacteria under flow we use a 3D Lagrangian tracker \cite{darnige2017lagrangian} which is based on real-time image processing, determining the displacement of a $xy$ mechanical stage to keep the chosen object at a fixed position in the observation frame. The z displacement is based on the refocusing of the fluorescent object, keeping the moving object in focus with a precision of a few microns in $z$. The acquisition frequency is 30~Hz. 
The Lagrangian tracker is composed of an inverted microscope (Zeiss-Observer, Z1) with a high magnification objective (100X/0.9 DIC Zeiss EC Epiplan-Neofluar), a $xy$ mechanically controllable stage with $z$ piezo-mover from Applied Scientific Instrumentation (ms-2000-flat-top-$xyz$) and a digital camera ANDOR iXon 897 EMCCD.
Trajectories are only considered when far away from the lateral walls (distances larger than 100 $\mu$m) and as long as they are within 5 $\mu$m from the surface. A typical error on this distance, resulting from the uncertainty on the $z$-detection as well as the uncertainty on the position of the bottom surface is around 3 $\mu$m.

\subsubsection{Fluorescence experiments}

For flagella visualization, we use a genetically modified strain from the AB1157 wild-type (AD1) published in Ref.~\cite{schwarz-linek2016escherichia}. This strain contains a FliC mutation to bind to the dye from Alexa Fluor.
Single colonies of frozen stocks are incubated overnight (≈16 h) in 5mL of liquid Luria Broth at 30 C, shaken for aeration at 200 rpm. The bacteria are washed and resuspended in Berg’s motility buffer (BMB: 6.2 mM K$_2$HPO$_4$, 3.8 mM KH$_2$PO$_4$, 67 mM NaCl, and 0.1 mM EDTA). For flagella staining, $0.5$mL of the bacterial suspension in BMB at $2 \times 10^9$ bact/mL are mixed with $5\mu$L of Alexa Fluor 546 C5-maleimide suspended at $5$mg/mL in DMSO. The sample is kept in the dark, shaking at $100$rpm for $1$h. Bacteria are then three times washed in BMB and finally suspended at $10^8$bact/mL in BMB supplemented with polyvinylpyrolidone (PVP 350 kDa: $0.005 \%$) to prevent sticking to the walls of the microchannel.
The solution is then seeded with passive particles to be used for flow velocity determination (latex beads from Beckman Coulter $d = 2 \mu$m, density $\rho = 1.027$g/mL at a volume fraction $\phi = 10^{-7}$).

The microfluidic PDMS channel is $H = 20 \mu$m deep, $W = 200 \mu$m wide and several millimeters long.
We capture the bacterial dynamics of bacteria within $5\mu$m from the surface, using an inverted microscope (Zeiss-Observer, Z1) with a high magnification objective (100 $\times$ /0.9 DIC Zeiss EC Epiplan-Neofluar) and a digital camera ANDOR iXon 897 EMCCD at $30$fps.
As bacteria are transported downstream, they are kept in the frame of observation by manually displacing the microscope's stage, which position is registered.
During post-processing we extract the bacterial positions and orientations from the images.

\subsubsection{Data analysis}

To determine the frequency of the bacterial oscillations, we Fourier transform the bacterial trajectories obtained from experiments for different shear rates. 
(1) In the case of the experiments using bacteria with fluorescently labelled flagella, the in-plane angle $\psi(t)$ of the cell orientation is determined by fitting an ellipse to the acquired camera image and distinguish between head and tail by the velocity director.
(2) In the case of the 3D tracking experiments, the orientation cannot be determined in the same manner, but the lateral velocity $v_z(t)$ is used to search for oscillatory motion.
Hence, either $\psi(t)$ or $v_z(t)$ are Fourier transformed for trajectories of sufficiently long duration to resolve the lowest and highest frequencies accurately. 
The frequency of each trajectory is determined by selecting the highest peak in the resulting Fourier spectrum.
This is repeated for all trajectories to form an ensemble of frequencies, from which we evaluate the mean frequency $f(\dot \gamma)$ and its standard deviation.

\subsection{Chirality-induced rheotaxis}
\label{subsec:RTFcalculation}

Marcos et al.\ used Resistive Force Theory (RFT) to calculate the rheotactic behaviour of helical bacteria in shear flows in the bulk \cite{Marcos2009,Marcos2012}.
Based on their work and using their \textsc{Mathematica} notebook, 
that includes the resistive force theory calculations for a helix subjected to shear flow,
we are able to identify the full angular dependence of the rheotactic torque [Eq. (\ref{Eq:OmH})].

In RFT a helical flagellum segment is approximated by a stiff slender rod with
anisotropic friction coefficients $\xi_\perp$ and $\xi_{||}$ with $1 < \xi_\perp / \xi_{||} < 2$. The viscous force per unit length opposing the motion of a rod is written as  $\vec{f} = -\xi_{||} \vec{u}_{||} - \xi_\perp  \vec{u}_\perp$, with the local rod velocity ($\vec{v}_l$) relative to the external shear flow ($\vec{v}_f=\dot{\gamma}y\hat{\vec{x}}$), $\vec{u} =\vec{v}_l -\vec{v}_f   = \vec{u}_{||} + \vec{u}_\perp$
where the local rod velocity $\vec{v}_l$ is a sum of its translational and rotational velocity  $\vec{v}_l = \vec{v} + \boldsymbol{\Omega} \times \vec{r}(s;\psi,\theta)$,
and $\vec{u}_{||}= (\vec{u}\cdot \hat{\vec{t}}) \hat{\vec{t}}$,
$\vec{u}_\perp = \vec{u} - \vec{u}_{||}$;
here $\vec{r}(s;\psi,\theta)$ is a space-curve of a helix parametrised by $s$ and oriented along the swimmer direction, given by the angles $\psi$ and $\theta$, and
the tangent is $\vec{t} = (d \vec{r}/ d s)/|d \vec{r}/ d s|$.
After integrating the force and torque on the helix at angles $\psi$ and $\theta$ over the full helix length and averaging over the helix phase angle, one can in principle solve for the unknown helix velocity $\vec{v}$ and angular velocity $\boldsymbol{\Omega}$.
While in a good approximation the helix will rotate in flow similar as a rigid rod-like particle \cite{Marcos2012}, the velocity $\vec{v}$ determines chirality-induced migration velocity.


The analytic expressions for  $\vec{v}$ obtained with \textsc{Mathematica} are rather lengthy and cannot be reduced or simplified by the program.
However, we can plot the velocity components $v_x$, $v_y$ and $v_z$ depending on its orientation angles $\psi$ and $\theta$ for a given helix shape. 
Fortunately, by trial and error, we could extract the angular dependencies of the velocities, giving
\begin{align}
v_x^H &= -k_1\dot{\gamma}\sin 2\theta\cos^2\psi\label{Eq:r1} , \\
v_y^H &= -k_1\dot{\gamma}\sin \theta \sin 2 \psi\label{Eq:r2} , \\
v_z^H &= 2k_1\dot{\gamma}(-\sin^2\theta\cos^2\psi + \cos 2\psi) 
\label{Eq:r3}
\end{align}
which linearly increase with the shear rate $\dot{\gamma}$, and
where the prefactor $k_1$ only depends on the helix geometry and is $k_1 > 0$ for a left-handed helix (as it is the case for the normal form of \textit{E.~coli} bacteria), and $< 0$ otherwise.

The swimmer is oriented along direction $\vec{e}=(-\cos\theta \cos\psi, \sin\theta, -\cos\theta \sin\psi)$, and a rheotactic torque can be expressed as
$\boldsymbol{\Omega}^H = -k_2\vec{e} \times \vec{v}$
where the prefactor $k_2$ depends on the shape of the cell body \cite{Marcos2012}.
The components $\Omega_x^H$, $\Omega_y^H$ and $\Omega_z^H$ are given by
\begin{align}
\Omega_x^H &=  -\bar{\nu}_{H}\dot{\gamma}(\cos 2\theta - 2\cos^2\theta\sin^2\psi), 
\label{Eq:r11}\\
\Omega_y^H &=  -\bar{\nu}_{H}\dot{\gamma}(\cos\theta+\cos 3 \theta)\cos\psi/2, 
\label{Eq:r12}\\
\Omega_z^H &= -\bar{\nu}_{H}\dot{\gamma}\cos^2\theta\sin\theta\sin 2 \psi, 
\label{Eq:r13}
\end{align}
 where $\bar{\nu}_{H}=2k_1k_2$.
The components of this torque in the $\psi$ and $\theta$ directions are written down in Eq. (\ref{Eq:OmH}).

\subsection{Simulations of surface rheotaxis}
\label{subsec:RheotaxisSimulations}

The bacterial surface rheotaxis is simulated by numerical integration of the orientation dynamics, encapsulated by the covariant Langevin equation \cite{raible2004langevin} written out in terms of the angles $(\psi, \theta)$ that live on the curved surface $|\vec{p}| =1$,
\begin{align}
\delta \psi
&= (\Omega^W_\psi +  \Omega^F_\psi +  \Omega^V_\psi)\delta t 
+ \frac{\sqrt{2 D_r \delta t}}{\cos \theta} ~ \eta_\psi,
\\
\delta \theta
&= (\Omega^W_\theta +  \Omega^F_\theta +  \Omega^V_\theta)\delta t 
- \tan \theta D_r \delta t
+ \sqrt{2 D_r \delta t} ~ \eta_\theta, \nonumber
\end{align}
where $D_r = 0.057 \mu \text{m}^2/\text{s}$ is the rotational diffusion coefficient \cite{Drescher2011}, the noise correlations are defined as
$\langle  \eta_i \rangle = 0$ and $\langle  \eta_i(t) \eta_j(t') \rangle = \delta_{ij} \delta (t-t')$ and
the deterministic terms are written out explicitly in equations (\ref{OmegaPsi}-\ref{OmegaTheta}) below.

At the start of a trajectory, the swimmer reaches the surface with a small negative pitch angle, $\theta(t=0)=-\pi/10$, and a random uniformly distributed in-plane angle, $\psi(0) \in [-\pi,\pi]$.
Subsequently, its orientation ($\psi(t)$, $\theta(t)$) is integrated using a forward Euler scheme with time step $\delta t = 10^{-3}$s, and after every time step the orientation angles are renormalised to their domains, $\psi \in [-\pi, \pi]$ and $\theta \in [-\pi/2, \pi/2]$, by setting 
$\psi \to \text{mod}(\psi, 2\pi, -\pi)$ and 
$\theta \to \text{abs}(\theta + \pi/2) - \pi/2$.
Next, the spatial dynamics is obtained by computing the velocity parallel to the wall, 
$\vec{v}_{||} = v_0 \vec{p}_{||} +  \dot{\gamma} \delta \hat{\vec{x}}$,
where $v_0=20\mu \text{m/s}$ is the swimming speed, $\vec{p}_{||} = p_x \hat{\vec{x}} +p_z \hat{\vec{z}} = (-\cos\theta\cos\psi, -\cos\theta\sin\psi)$ is the swimmer orientation parallel to the wall, and $\delta=W/2=0.5 \mu$m is the distance from the wall.
The surface trajectories are then found by numerical integration of $\dot{\vec{r}}_{||} = \vec{v}_{||}$.
A trajectory ends when the pitch angle exceeds the escape angle, $\theta_e = \pi/6$, after which the swimmer escapes the surface.

Hence, $N = 10^4$ trajectories are simulated for 10 values of the applied shear rate, $\dot \gamma = 10^{2.5 (i-1)/9}$, where $i=1,2, \dots, 9,10$. 
From these trajectories $(\psi, \theta)$ the distribution of the in-plane angle, PDF($\psi$), follows immediately.
To determine the frequency of the bacterial oscillations, a trajectory must be sufficiently long to resolve the smaller frequencies.
At high shear the average trajectory (residence) time is smaller than at low shear, so we discard trajectories shorter than $10 \text{s}$, which leaves at least $\sim 100$ trajectories for any shear rate.
The frequencies of the remaining trajectories are then obtained individually by Fourier transforming the in-plane angle $\psi(t)$, and selecting the frequency of the highest peak in the resulting Fourier spectrum.
Using this ensemble, we evaluate the mean frequency $f(\dot \gamma)$ and its standard deviation.

\subsection{Estimation of model parameters}
\label{subsec:EstimateParameters}

In the following we present the parameters we used to produce Figs.~\ref{Fig:3} and \ref{Fig:4} above.

\textit{Swimmer aspect ratio $\Gamma$:} While cell bodies of \textit{E.~coli} bacteria have typical aspect ratios of $\approx 3$, including the flagella bundle increases the effective aspect ratio $\Gamma$ and we choose $\Gamma=5$ for convenience.
Note that $\Gamma$ does not enter the model directly, but rather $G= \frac{\Gamma^2-1}{\Gamma^2+1} \lesssim 1$, which does not change significantly with $\Gamma$.

\textit{Swimming speed $v_s$:} 
Because we use non-tumbling swimmers in our experiments we choose a constant swimming speed $v_s = 20\mu$m/s.

\textit{Hydrodynamic/steric reorientation frequency at wall $\nu_W$:} Steric reorientation rates have been reported to be of the order $\nu_W^{steric} \sim 1-10s^{-1}$ for flagellated \textit{Caulobacter} bacteria \cite{Li2009}.
Reorientation rates away from the walls due to hydrodynamic interactions can be approximated by far-field expressions \cite{Spagnolie2012}.
with a prefactor $\nu_W^{Hi} = 3p/(128\pi\eta h^3)$, where $p=0.8pN\,\mu$m is the dipole strength of \textit{E.~coli} bacteria \cite{Drescher2011}, $\eta=10^{-3}Pa\,s$ the viscosity of water, and $h$ the distance of the swimmer from the surface.
If far-field hydrodynamics would still hold close to the wall ($h \approx 1-2\mu$m), one would find $\nu_W^{HI}\approx 0.75-6 s^{-1}$, which is smaller than the steric contribution but also acts when the swimmer moves away from the surface, so it could increase the wall residence time \cite{Schaar2015}. Also note that the $1/h^3$  dependence likely gives an overestimate when a swimmer is very close to the wall, $h <1\mu$m, as the multipole approximation breaks down. 
Taking this information together, we use $\nu_W=3s^{-1}$ to capture the combined effects of steric and hydrodynamic interactions.

\textit{Downstream advection:} 
To calculate the downstream advection velocity due to the linear shear flow, we use the distance from the surface $h_s = 1\mu$m, so the velocity is $\vec{v}_f = \dot{\gamma} h_s \hat{\vec{x}}$.

\textit{Equilibrium pitch angle $\theta_0$:}
Recent results have shown that the equilibrium pitch angle $\theta_0$ is not exactly parallel to the wall but sightly points into the surface \cite{Bianchi2017}.
We have included this in our model by considering the surface alignment term with the equilibrium angle dependence $\sin (2 (\theta-\theta_0))$, but this did not change our results much. 
Therefore we omit it for clarity, setting $\theta_0=0$. 
However, we still allow explicitly for dynamics in $\theta$ from the other contributions.

\textit{Circling frequency near a wall $\nu_C$:} The typical circling frequencies of an \textit{E.~coli} bacterium close to a boundary is on the order of $\sim 1$s \cite{Luzio2005, Lauga2006}. Since the frequency $\Omega^W_\psi \approx 2 \nu_C$ for $G\sim1$, we choose $\nu_c=0.5s^{-1}$, which gives typical circles of radius $R=v_0 / \Omega^W_\psi \sim 20 \mu$m. Note that there is quite a variety in these frequencies between different individual bacteria \cite{Lauga2006,Li2008}, which will lead to bacteria having different individual critical shear rates. 

\textit{Chirality-induced bulk reorientation rate $\bar{\nu}_{H}$:}
The rheotactic drift for bacteria in bulk has been quantified by Marcos et al. \cite{Marcos2009,Marcos2012}, as described in \S II above. We expect similar rheotactic strength for our bacteria, which is satisfied approximately for $\bar{\nu}_{H}=0.02$.

\textit{Anchoring reorientation rate $\bar{\nu}_{V}$}:  Due to a similar size of the head, we expect that the order of magnitude for a bacterium is comparable to the values found for sperm cells \cite{Tung2015}, and we chose $\bar{\nu}_{V}=0.75$.
Note, the magnitude of this effect is reduced strongly when the swimmer is not in close proximity to the walls, since it relies on enhanced friction obtained from the lubrication regime \cite{Tung2015, Daddi2018state}. When the swimmer is oriented away from the surface ($\theta > 0$) we expect this effect to vanish quickly, quantified by a small value $\theta_V=0.04$ used in the tanh-function [see Eq.~(6)].

\textit{Escape angle $\theta_e$:} In order to determine when a swimmer leaves a surface, it has to reach a certain escape angle $\theta_e$ which we chose to be $\theta_e=\pi/6$, following references \cite{Drescher2011, Schaar2015, Mathijssen2016b}. Note that $\theta_e$ does not influence the dynamics in the model per se, but rather defines how long a bacterium stays at a surface.

\textit{Rotational diffusion:} 
The orientation of the bacteria is affected by fluctuations as they swim. We use the rotational diffusion coefficient $D_r = 0.057 \mu \text{m}^2/\text{s}$ \cite{Drescher2011}. 

\textit{Tumbling:} 
We do not include tumbling in our simulations presented in the main text because we use smooth swimmers in our experiments, but in Supplementary Fig.~S2 we show orientation distributions for simulations that include tumbles. We simulate these tumbling events by temporarily increasing the swimmer's rotational diffusion coefficient \cite{Mathijssen2016b}, to $D_T = \varphi^2 / (2 \tau_T)$, where the average tumble angle $\varphi =\pi/3 = 60$ degrees \cite{berg1972chemotaxis}, the tumbling time is $\tau_T = 0.1$s, and the duration between tumbles is exponentially distributed with average run time $\tau_R = 1$s.

\subsection{Model fixed points and frequencies}
\label{subsec:ModelFixedPoints}

\subsubsection{Summary of equations}

Combining the deterministic contributions of our rheotactic model, we have
\begin{align}
\label{OmegaPsi}
\Omega_\psi &=
\nu_{C} (1-3\sin^2\theta + G \cos^2\theta(1+3\sin^2\theta) \nonumber \\ &
+ \frac{\dot{\gamma} \nu_J}{2} (1+G) \sin\psi\tan\theta  \nonumber \\ &
+  \dot{\gamma} \bar{\nu}_{H} \cos\psi \frac{\cos 2 \theta }{\cos\theta}  
- \dot{\gamma} \bar{\nu}_{V} \sin \psi  \frac{1 - \tanh \frac{\theta}{\theta_V}}{2},
\end{align}
\begin{align}
\label{OmegaTheta}
\Omega_\theta &=
- \nu_{W}\sin 2 \theta \left( 1+ \frac{G}{2} \left(1+ \cos^2 \theta \right) \right)  \nonumber \\ &
+ \frac{\dot{\gamma} \nu_J}{2} \dot{\gamma} (1 - G \cos 2 \theta) \cos\psi  \nonumber \\ &
+  \dot{\gamma} \bar{\nu}_{H} \sin\psi\sin \theta  
- \dot{\gamma} \bar{\nu}_{V} \sin \theta  \frac{1 - \tanh \frac{\theta}{\theta_V}}{2}.
\end{align}

\subsubsection{Adler transition from circling to straight motion}

The first transition, from circling to straight motion, has previously been described in the literature for sperm cells \cite{Tung2015}.
It can be characterised as the point at which the upstream-directed torque from weathervane effect becomes more important than the constant torque from the bacterial circling on surfaces.
This can be captured in the limit of small pitch angles, $\theta \to 0$, where the one-dimensional equation for the in-plane angle (\ref{OmegaPsi}) simplifies to
\begin{align}
\Omega_\psi &=
\nu_{C} (1 + G) - \frac 1 2 \dot{\gamma} \bar{\nu}_{V} \sin \psi + \dot{\gamma} \bar{\nu}_{H} \cos \psi.
\end{align}
This can immediately be solved for the equilibrium angle  $\psi_0$ where $\Omega_\psi = 0$.
The solution is a little long to write down, but is nothing more than an arctangent.
Further progress can be made by noting that $\bar{\nu}_{H} \ll \bar{\nu}_{V}$ and also $\cos \psi \ll  \sin \psi$, near $\psi \sim \pi/2$, at shear rates close to the transition where the bacteria swim to the right.
Consequently, the resulting equilibrium angle is given by
\begin{align}
\label{AdlerPsi}
\psi_0 &= \arcsin \frac{2(1+G) \nu_C}{\bar{\nu}_{V} \dot{\gamma}}.
\end{align}
This function has no real solutions for shear rates smaller than a critical value.
Indeed, the rotating bacteria do not have a stable equilibrium orientation, but for large enough shear they break out of their circles and maintain a constant bearing.
Therefore, we found the critical shear rate of Adler transition,
\begin{align}
\label{eq:FirstCriticalShear}
\dot{\gamma}_{c1}^\text{th} &= \frac{2(1+G) \nu_C}{\bar{\nu}_{V} }.
\end{align}

\subsubsection{Approximation of the equilibrium orientations}

In order to estimate the equilibrium rheotactic orientations at higher shear rates above the Adler transition, the one-dimensional approach breaks down because the pitch angle $\theta$ becomes significant.
This shortcoming is also observed from the 1D solution (\ref{AdlerPsi}) that only decreases with increasing shear but does not capture the later increase, which gives rise to an optimum shear rate for upstream swimming.

Therefore, we aim to solve for both $\psi_0$ and $\theta_0$ such that $\Omega_\psi = \Omega_\theta = 0$.
Because it is known that the bacteria swim to the right and left at high shear rates,
we linearise equations (\ref{OmegaPsi}, \ref{OmegaTheta}) about these directions and also consider small pitch angles parallel to the surface:
\begin{align}
\label{linearisation}
(\psi, \theta) &= \left ( \alpha \pm \frac{\pi}{2}, \beta \right),
\end{align}
where $\alpha, \beta \ll 1$ and where the top sign ($+$) corresponds to swimming to the right and the bottom sign ($-$) corresponds to swimming to the left.
To first order in $\alpha$ and $\beta$, that yields the linear expression
\begin{align}
\left[
{\begin{array}{c}
   \Omega_\psi \\
   \Omega_\theta \\
  \end{array} }
\right]
&=
\left[
{\begin{array}{c}
E \\ 
0 \\
\end{array} }
\right]
+
\left[
{\begin{array}{cc}
A & B \\
C & D
\end{array} }
\right]
\left[
{\begin{array}{c}
\alpha \\ \beta
\end{array} }
\right],
\end{align}
where the shear-dependent coefficients are
\begin{align}
\label{Evalue}
A &=  \mp \bar{\nu}_{H} \dot{\gamma}, \quad
B = \pm \frac{1}{2} \nu_J \dot{\gamma} (1+G) \pm \frac{ \bar{\nu}_{V} \dot{\gamma}}{2 \theta_V}, \\
C &= \mp \frac{1}{2} \nu_J \dot{\gamma} (1-G) , \quad
E = \nu_W (1+G) \mp \frac{1}{2} \bar{\nu}_{V} \dot{\gamma}, \nonumber \\
D &= -2 \nu_W (1+G) - \frac{1}{2} \bar{\nu}_{V} \dot{\gamma} \pm \bar{\nu}_{H} \dot{\gamma}. \nonumber
\end{align}
This matrix equation can be inverted directly, giving the equilibrium in-plane and pitch angles,
\begin{align}
\label{EqiPsi}
\psi_0 = \frac{D E}{B C - A D} \pm \frac{\pi}{2},
\quad \quad
\theta_0 = \frac{C E}{A D - B C}.
\end{align}
These results are shown as the dotted lines in Fig.~\ref{Fig:4}(a).

\subsubsection{Optimal shear rate for upstream swimming}

Continuing from the above theory, it is also possible to extract an estimate for the optimal shear rate for upstream swimming.
This occurs when the right equilibrium in-plane angle (given by Eqn.~\ref{EqiPsi} with the top of the $\pm$ signs, and the green lines in Fig.~4b) is closest to the upsteam orientation ($\psi \to 0$), thus at the minimum with respect to the shear rate.
Hence, we seek to solve for
\begin{align}
\label{eq:OptimalUpstreamShear}
0 = \frac{\partial}{\partial \dot{\gamma}} \left[
\frac{(d_0 + d_1 \dot{\gamma}) (e_0 + e_1 \dot{\gamma})}{ ( b_1 \dot{\gamma}) (c_1 \dot{\gamma}) - (a_1 \dot{\gamma}) (d_0 + d_1 \dot{\gamma})}
\right],
\end{align}
where we wrote $D = d_0 + d_1 \dot{\gamma}$, and similar for the other functions. This readily gives the desired optimum, the best shear rate for upstream swimming, $\dot{\gamma}_{u}^\text{th} \approx 5.95 \persec$.
Inserting this value into  Eqn.~\ref{EqiPsi} then also gives the most upstream orientation, $\psi_{u}^\text{th} \approx 26.7$ degrees.

\subsubsection{Oscillation frequency}

We start with a linear stability analysis of the equilibrium orientation angles,
\begin{align}
(\psi, \theta) &= \left (\psi_0 + \epsilon_\psi , \theta_0 + \epsilon_\theta \right),
\end{align}
where $\epsilon_\psi, \epsilon_\theta \ll 1$. 
Expanding the equations of motion to first order in $\epsilon_\psi$ and $\epsilon_\theta$ again yields
\begin{align}
\left[
{\begin{array}{c}
   \Omega_\psi \\
   \Omega_\theta \\
  \end{array} }
\right]
& \approx
\left[
{\begin{array}{cc}
A & B \\
C & D 
\end{array} }
\right]
\left[
{\begin{array}{c}
  \epsilon_\psi \\ \epsilon_\theta
  \end{array} }
\right],
\end{align}
in terms of $(A,B,C,D)$ given by Eqs.~(\ref{Evalue}). The eigenvalues of this matrix are
\begin{align}
\label{eigenvalues}
\lambda &= \frac 1 2 \left(
A + D \pm \sqrt{(A-D)^2 + 4 B C}
\right)
\end{align}
The real part of the eigenvalues characterises the stability of the equilibrium orientation.
For our model parameters it is always negative for all shear rates, 
indicating a stable equilibrium because of the surface alignment ($\nu_W$ term).

The imaginary part, however, characterises the presence of oscillations.
For small shear rates the imaginary part is zero, but above a certain shear rate oscillations emerge. 
This occurs when $(A-D)^2 = 4 B C$, at the critical shear rate
\begin{align}
\label{eq:SecondCriticalShear}
\dot{\gamma}_{c2}^\text{th} =
\frac{4(1+G)  \nu_W}{2 \sqrt{(1-G^2) + (1-G) \bar{\nu}_{V} / \theta_V} - \bar{\nu}_{V}  \mp 4 \bar{\nu}_{H} }.
\end{align}
Below this shear rate value, the equilibrium orientation is a ``star-type'' stable fixed point, whereas above $\dot{\gamma}_O$  it is a ``spiral-type'' stable fixed point with damped oscillations. 

The oscillation frequency is given directly by the imaginary part of the eigenvalues (Eq.~\ref{eigenvalues}), shown as the green line in Fig.~\ref{Fig:4}(c). For high shear rates, this tends to the linear function for the oscillation frequency
\begin{align}
\label{eq:OscillationFrequencyTheory}
\omega_O =
\frac{\dot{\gamma}}{4}  \sqrt{4(1-G^2) +4(1-G) \frac{\bar{\nu}_{V}}{\theta_V} - \bar{\nu}_{V}^2 \pm 8 \bar{\nu}_{H} \bar{\nu}_{V} - 16 \bar{\nu}_{H}^2 }.
\end{align}
If the weathervane and helix coefficients are small, this simplifies to the pure Jeffery frequency
\begin{align}
\label{eq:OscillationFrequencyJeffery}
\omega_O =
\frac{1}{2} \dot{\gamma} \sqrt{(1-G^2)}.
\end{align}

\normalsize 

\section*{Acknowledgements}
We are grateful to the authors of Ref.~\cite{Marcos2012}, and in particular to Henry Fu, providing the \textsc{Mathematica} notebook including the resistive force theory calculations for a helix subjected to shear flow. We thank Angela Dawson and Vincent Martinez for providing the AB1157 \textit{E. coli} strain. 
AM acknowledges funding from the Human Frontier Science Program (Fellowship LT001670/2017).
EC, NFM, AL, GM acknowledge funding from the ANR-15-CE30-0013 BacFlow.
AL and NFM acknowledge funding from the ERC Consolidator Grant PaDyFlow (Agreement 682367).
AZ acknowledges funding from the European Union's Horizon 2020 research and innovation programme under the Marie Sk\l{}odowska-Curie grant (Agreement 653284).

\section*{Author contributions}
AM, NFM, GJ, EC, AL and AZ designed the research, NFM and GJ performed the experiments, AM, NFM and GJ analysed the experimental data, AM and AZ developed the theory and simulations, AM, NFM, AL and AZ wrote the manuscript.

\section*{Competing interests}
The authors declare no competing financial interests.

\bibliography{biblio_surface_rheotaxie}

\end{document}